\documentclass[runningheads,a4paper]{llncs}

\usepackage{amsmath,amssymb}
\usepackage{graphicx}

\usepackage{eventB}
\usepackage{subfig}
\usepackage{zed-csp}
\usepackage{verbatim}
\usepackage{graphicx}

\usepackage{url}
\urldef{\mailsa}\path|{alfred.hofmann, ursula.barth, ingrid.haas, frank.holzwarth,|
\urldef{\mailsb}\path|anna.kramer, leonie.kunz, christine.reiss, nicole.sator,|
\urldef{\mailsc}\path|erika.siebert-cole, peter.strasser, lncs}@springer.com|

\usepackage{tikz}
\usetikzlibrary{arrows}

\def\setminus{-}

\newcommand{\Bmachine}{\;\textbf{machine}\;}

\newcommand{\Bconstants}{\;\textbf{constants}\;}

\newcommand{\Bproperties}{\;\textbf{properties}\;}

\newcommand{\Bvariables}{\;\textbf{variables}\;}

\newcommand{\Binvariant}{\;\textbf{invariant}\;}
\newcommand{\Bvariant}{\;\textbf{variant}\;}

\newcommand{\Bevents}{\;\textbf{events}\;}

\def\ord{{\tt o}}
\def\anticipated{{\tt a}}
\def\convergent{{\tt c}}

\def\Bany{\;\mbox{\bf any}\;}
\def\Bwhere{\;\mbox{\bf where}\;}
\def\Bwhen{\;\mbox{\bf when}\;}
\def\Bthen{\;\mbox{\bf then}\;}
\def\Bend{\;\mbox{\bf end}}
\def\Brefines{\;\mbox{\bf refines}\;}
\def\Bstatus{\;\mbox{\bf status}}
\def\Bconvergent{\;\mbox{convergent}}
\def\Banticipated{\;\mbox{anticipated}}

\def\Bordinary{\;\mbox{ordinary}}

\def\comp{;}

\newcommand{\eventref}{\preccurlyeq}

\newcommand{\Weventref}{\preccurlyeq_W}

\begin{document}

\mainmatter  

\title{Managing LTL properties in Event-B refinement}

\titlerunning{Managing LTL properties in Event-B refinement}

%
%
\author{Steve Schneider\inst{1} \and
Helen Treharne\inst{1} \and \\Heike Wehrheim\inst{2}
\and David M. Williams \inst{3}} 
\authorrunning{Schneider, Treharne, Wehrheim, Williams}


\institute{University of Surrey, England, UK \and University of Paderborn, Germany \and VU University Amsterdam}

%
%

\toctitle{}
\tocauthor{}
\maketitle

\begin{abstract}
%
%

Refinement in Event-B supports the development of systems via proof based step-wise refinement of events.   This refinement approach ensures safety properties are preserved, but additional reasoning is required in order to establish liveness and fairness properties.
In this paper we present results which allow a closer integration of two formal methods, Event-B and linear temporal logic.   In particular we show how a class of temporal logic properties can carry through a refinement chain of machines. Refinement steps can include introduction of new events, event renaming and event splitting. We also identify a general liveness property that holds for the events of the initial system of a refinement chain.  The approach will aid developers in enabling them to verify linear temporal logic properties at early stages of a development, knowing they will be preserved at later stages.  We illustrate the results via a simple case study.


\end{abstract}

\section{Introduction}
\label{sec:intro}
%
%
%
%

Event-B~\cite{Abrial09} is a step-wise development method with excellent tools: Rodin platform~\cite{DBLP:journals/sttt/AbrialBHHMV10} providing proof support and ProB~\cite{DBLP:journals/sttt/LeuschelB08} providing model checking. As Hoang and Abrial~\cite{DBLP:conf/icfem/HoangA11} clearly state the focus of verification within Event-B has been on the safety properties of a system to ensure that ``something (bad) never happens". Typically, this has been done via the discharging of proof obligations. Nonetheless, the use of linear temporal logic (LTL) to specify temporal liveness properties has also been prevelant, for example in its application within the ProB tool~\cite{DBLP:conf/fm/LeuschelFFP09}.   The challenge is to identify more natural ways of integrating Event-B and LTL, so that LTL properties can be preserved by Event-B refinement, which is not currently the case in general.



Event-B describes systems in terms of {\em machines} with state, and {\em events} which are used to update the state.  Events also have {\em guards}, which are conditions for the event to be enabled.  One (abstract) machine may be refined by another (concrete) machine, using a {\em refinement step}.  A {\em linking invariant} captures how the abstract and concrete states are related, and each abstract event must be refined by one or more concrete events whose state transformations match the abstract one in the sense of preserving the linking invariant.  Refinement is transitive, so a sequence of refinement steps, known as a {\em refinement chain}, will result in a concrete machine which is a refinement of the original abstract one.  

A particular feature provided by Event-B is the introduction of {\em new} events in a refinement step---events which do not refine any abstract event.  This allows for refinements to add finer levels of granularity and concretisation as the design develops; there are many examples in~\cite{Abrial09}.  These new events are invisible at the abstract level (they correspond to the abstract state not changing), and we generally need to verify that they cannot occur forever.  Event-B makes use of {\em labels} to keep track of the status of events as a refinement chain  progresses.   Event-B labels are {\it anticipated}, {\it convergent} and {\it ordinary}.    The labelling of events in Event-B form part of the core of a system description but their inclusion is primarily to support the proof of safety properties and ensuring that events cannot occur forever: convergent events must decrease a variant and anticipated events cannot increase it.  In this paper all newly introduced events must be convergent or anticipated, and anticipated events must become convergent at some stage. 
As an initial example, consider a $Lift$ machine with two events {\em top} and {\em ground}, representing movement to the top and to the ground floor.   This can be refined by a machine $Lift'$ introducing two new anticipated events {\em openDoors} and {\em closeDoors}.   The events {\em top} and {\em ground} are blocked when the doors are open, but enabled when the doors are closed.

Linear temporal logic provides a specification language for capturing properties of executions of systems, and is appropriate for reasoning about liveness and fairness.  For example, we might verify for $Lift$ that whenever {\em top} occurs, then eventually {\em ground} will occur.  However, this is not guaranteed for its refinement $Lift'$: it may be that the doors open and close repeatedly forever following the {\em top} event, thus never reaching the next {\em ground} event.  Alternatively it may be that the system deadlocks with the doors open, again preventing {\em ground} from occurring.   Hence we see that LTL properties are not automatically preserved by Event-B refinement.    In the first case we would require some assurance that {\em openDoors} and {\em closeDoors} cannot repeat forever without the lift moving; in the second case we require some liveness property on {\em closeDoors} to prevent termination with the doors open.

In this paper we present results for when temporal logic properties can be carried through Event-B refinement chains. The results generalise to events that are split---refined by several events---during a refinement chain.   We also identify conditions on temporal logic properties that make them suitable for use in a refinement chain, since some properties are not preserved by Event-B refinement (for example, the property ``{\em closeDoor} never occurs'' holds for $Lift$ but not for its refinement $Lift'$).  
The results are underpinned by our process algebra understanding of the Event-B semantics, in particular the traces, divergences and infinite traces semantics used for CSP and applied to Event-B in \cite{DBLP:journals/fac/SchneiderTW14}.

The paper is organised as follows: Section~\ref{sec:eventb} provides the necessary Event-B refinement background and the refinement strategy we use in the paper. Section~\ref{sec:example} introduces a running example. Section~\ref{sec:ltl} defines the LTL we use.  Sections~\ref{sec:preserving} and~\ref{sec:extendingPreserving}
present and illustrate the main theoretical results.  Proofs are included in the Appendix.
We put our work into the context of related work in Section~\ref{sec:related_work} and our future work in Section~\ref{sec:future_work}.

\section{Event-B}
\label{sec:eventb}

%

\subsection{Event-B Machines}
\label{sec:event_b_machines}

An Event-B development is defined using {\it machine}s. A machine $M$ contains a vector of variables and a set of events. The {\em alphabet} of $M$, $\alpha M$, is the set of events defined in $M$. Each event $evt_i$ has the general form $evt_i \defs \Bany x \Bwhere $$G_i(x,v)$ $\Bthen $$ v :| BA_i(v,x,v') \Bend$, where $x$ represents the parameters of the event, the guard $G_i(x,v)$ is the condition for the event to be enabled.  The body is given by $v :| BA_i(v,x,v')$ whose execution assigns to $v$ any value $v'$ which makes the {\it before-after} predicate $BA_i(v,x,v')$ true. This simplifies to $evt_i \defs \Bwhen G_i(v)$  $\Bthen $$ v :| BA_i(v,v') \Bend$ when there are no parameters, since the guard and the  {\it before-after} predicate does not refer to the parameters $x$.

Variables of a machine are initialised in an initialisation event {\it init} and are constrained by an invariant $I(v)$. The Event-B approach to semantics is to associate proof obligations with machines. The key proof obligation, {\bf \tt INV}, is that all events must preserve the invariant. There is also a proof obligation on a machine with respect to deadlock freedom which means that a guard of at least one event in $M$ is always enabled.  When this obligation holds $M$ is {\it deadlock free}.

\subsection{Event-B Refinement}
\label{sec:event_b_refinement}
An Event-B development is a sequence of B machines linked by a refinement relationship. In this paper we use $M$ and $M'$ when referring to a refinement between an {\it abstract} machine $M$ and a {\it concrete} machine $M'$ whereas a chain of refinements is referred to using numbered subscripts, i.e., $M_0$, $M_i$, $\ldots$, $M_n$, to represent the specific refinement levels.

A refinement machine can introduce new events and split existing events. We omit the treatment of merging events in this paper. New events are treated as refinements of {\it skip}, i.e., $evt'_i$ does not refine an event in $M$. Note that when splitting events, $M'$ has several events $evt'_i$ refining a single event $evt_i$.

A machine $M$ is considered to be refined by $M'$ if the given {\it linking invariant} $J$ on the variables between the two machines is established by their initialisation, and preserved by all events. This  requirement is captured by the {\bf \tt INV\_REF} proof obligation. Formally, we denote the refinement relation between two machines, written $M \eventref M'$, when all the following proof obligations hold: feasibility {\bf \tt FIS\_REF}, guard strengthening {\bf \tt GRD\_REF} and simulation {\bf \tt INV\_REF}. Feasibility of an event is the property that, if the event is enabled (i.e., the guard is true), then there is some after-state. Guard strengthening requires that when a concrete event is enabled, then so is the abstract one. Finally, simulation ensure the occurrence of events in the concrete machine can be matched in the abstract one (including the initialization event). Further details of these proof obligations can be found in~\cite{Abrial09}. 

In Section~\ref{sec:intro} we introduced the three kinds of labelling of events in Event-B: {\it anticipated} ({\tt a}), {\it convergent} ({\tt c}) and {\it ordinary} ({\tt o}) and noted that convergent events are those which must not execute forever whereas {\it anticipated} events provide a means of deferring consideration of divergence-freedom until later refinement steps. The proof obligation that deals with divergences is  {\bf \tt WFD\_REF}. It requires that the proposed variant $V$  of a refinement machine satisfies the appropriate properties: that it is a natural number, that  it decreases on occurrence of any convergent event, and that it does not increase on occurrence of any anticipated event. Therefore, we augment the previous refinement relation with {\bf \tt WFD\_REF} such that $M \Weventref M'$. Ordinary events can occur forever and therefore {\bf \tt WFD\_REF} is not applicable for such events.


\subsection{Event-B Development strategies}
\label{sec:strategies}

\label{subsec:dev}

Event-B has a strong but flexible refinement strategy which is described in~\cite{Hallerstede2013272}. In~\cite{DBLP:journals/fac/SchneiderTW14} we also discussed different Event-B refinement strategies and characterised them with respect to the approaches documented by Abrial in~\cite{Abrial09} and supported by the Rodin tool.  
In this paper we focus on the simplest strategy, and the one most commonly used. The strategy has the following set of restrictions on a refinement chain $M_0 \Weventref M_1 \Weventref \ldots \Weventref M_n$:
\begin{enumerate}
\item all events in $M_0$ are labelled ordinary. This set of events is referred to as $O_0$.
\item each event of $M_i$ is refined by at least one event of $M_{i+1}$;
\item each new event in $M_{i}$  is either anticipated or convergent, where $i >0$;
\item each event in $M_{i+1}$ which refines an anticipated event of $M_i$ is itself either convergent or anticipated;
\item refinements of convergent or ordinary events of $M_i$ are ordinary in $M_{i+1}$.
\item no anticipated events remain in the final machine.  
\end{enumerate}

Figure~\ref{fig:strategy} illustrates our development strategy for a vending machine, detailed in Section~\ref{sec:example}, where $C_i$ is the set of convergent events within $M_i$, and $O_i$ is the set of ordinary events within $M_i$.  

For example, $O_0 = \{selectBiscuit, selectChoc, dispenseBiscuit, dispenseChoc\}$ and $C_0 = \emptyset$ in $VM_1$. In $VM_2$ we note that $C_1 = \{refund\}$. In $VM_3$ we note that $C_2 = \{refill\} $ and in $VM_4$ we have $C_3 = \{pay\}$. Thus we denote $C_{all} = C_1 \cup C_2 \cup C_3$. 

%
%

\begin{figure}[t]
\begin{centering}
\begin{scriptsize}
\begin{tabbing}
\hspace*{0.4cm}$selectBiscuit$ (\ord) \hspace*{0.4cm} \=$selectBiscuit$ (\ord)\hspace*{0.5cm} \=$selectBiscuit$ (\ord)\hspace*{0.5cm} \=$selectBiscuit$ (\ord)\\
\hspace*{0.4cm}$selectChoc$ (\ord) \>$selectChoc$ (\ord)\>$selectChoc$ (\ord)\>$selectChoc$ (\ord)\\
\hspace*{0.4cm}$dispenseBiscuit$ (\ord) \>$dispenseBiscuit$ (\ord)\>$dispenseBiscuit$ (\ord)\>$dispenseBiscuit$ (\ord)\\
\hspace*{0.4cm}$dispenseChoc$ (\ord) \>$dispenseChoc$ (\ord)\>$dispenseChoc$ (\ord)\>$dispenseChoc$ (\ord)\\
\hspace*{0.4cm}\>$pay$ (\anticipated)\>$pay$ (\anticipated)\>$pay$ (\convergent)\\
\hspace*{0.4cm}\>$refund$ (\convergent)\>$refund$ (\ord)\>$refund$ (\ord)\\
\hspace*{0.4cm}\>\>$refill$ (\convergent)\>$refill$ (\ord)\\\\
\hspace*{0.4cm}$VM_1$ \>$VM_2$ \>$VM_3$ \>$VM_4$
\end{tabbing}
\end{scriptsize}
\end{centering}
\caption{Events and their annotations in the Vending Machine development
\label{fig:strategy}
}
\end{figure}

\subsection{Event-B Semantics}
\label{sec:event-B_semantics}
 In this paper we define a trace of $M$ to be either an infinite sequences of events (\anticipated,\convergent \;or \ord), i.e.,  $\trace{e_0,e_1, \ldots}$ or a finite sequence of events, i.e., $\trace{e_0, \ldots, e_{k-1}}$ where the machine $M$ deadlocks after the occurrence of the final event. Traces correspond to maximal executions of machines. Plagge and Leuschel in~\cite{prob_ltl} provided a definition of an infinite or finite path $\pi$ of $M$ in terms of a sequence of events and their intermediate states. In order to distinguish notation we use $u$ to represent a trace without the intermediate states.  We need not consider the particular states within a trace in our reasoning which is based on infinite traces. When a machine $M$ is deadlock free all of its traces are infinite. We use the functions of concatenation ($\cat$), projection ($\project$) and length ($\#$) on finite and infinite traces.


A more complex behavioural semantics for B machines was given by Schneider {\it et al.} in~\cite{DBLP:journals/fac/SchneiderTW14} based on the weakest precondition semantics of \cite{morgan,butlerphd} for action systems and CSP. In~\cite{DBLP:journals/fac/SchneiderTW14} there are two key results that enable us to reason about infinite sequences of convergent and ordinary events in this paper. Firstly, the following predicate captures that if an infinite trace $u$ performs infinitely many events from $C$ then it has infinitely many events from $O$, where $C$ and $O$ are sets of events. 
\begin{definition}
$CA(C,O)(u) \defs (\# (u \project C) = \infty \implies \# (u \project O) = \infty)$
\end{definition}
$C$ and $O$ will be used to capture convergent and ordinary events through a development.
For an Event-B machine $M$ the above means that it {\em does not diverge on its $C$ events}.  This is precisely what we get when we prove {\bf \tt WFD\_REF} but the above definition describes the result on traces.

The second result from~\cite{DBLP:journals/fac/SchneiderTW14}, restated as Theorem~\ref{thm:chain_ca}, allows us to conclude that there are no infinite sequences of convergent events in the final machine of a refinement chain $M_n$. The function $g_{1,n}$ defines a compositional mapping for all concrete events to abstract events in terms of a function mapping $f$ at each refinement level  where $f_{i+1} : \alpha M_{i+1} \psurj \alpha M_i $ and  $f_{i+1}(evt_{i+1}) = evt_i$  $\Leftrightarrow$ $evt_{i+1}\; \mbox{\bf refines}\; evt_i$.  (Note that $g_{1,0}$ is the identity function.) 

\begin{definition}
\label{defn:g}
$g_{i,j}  =  f_j \comp f_{j-1} \comp \ldots \comp f_i$
\end{definition}

\begin{theorem} 
\label{thm:chain_ca}
If
$M_0 \Weventref M_1 \Weventref \ldots \Weventref M_n$
then 
\begin{eqnarray*}
M_n & \sat & CA(g_{1,n}^{-1}(C_0) \cup \ldots \cup g_{i,n}^{-1}(C_i) \cup \ldots \cup  C_n  \;,\; g_{1,n}^{-1}(O_0))
\end{eqnarray*}
\end{theorem}

\noindent The result for our example is simply $VM_4 \sat CA(C_{all},O_0)$ since there is no renaming: each function mapping $f_i$  is the identity.


\section{Example}
\label{sec:example}

\begin{figure}[th!]
\begin{center} 
\framebox{
$\begin{array}{l}
\Bmachine VM_1 \\
\Bvariables chosen\\ 
\Binvariant  chosen \subseteq\{ choc, biscuit \} \\
\Bevents \\
\quad \mbox{init} \defs 
 chosen := \{ \} \\ 
 \quad \mbox{selectBiscuit} \defs  \Bstatus: \Bordinary \\
 \qquad \Bwhen biscuit \not \in chosen 
\; \Bthen chosen := chosen \union \{ biscuit \} 
    \Bend \\ 
 \quad \mbox{selectChoc} \defs  \Bstatus: \Bordinary \\
 \qquad \Bwhen choc \not \in chosen 
\; \Bthen chosen := chosen \union \{ choc \} 
    \Bend \\ 
 \quad \mbox{dispenseBiscuit} \defs  \Bstatus: \Bordinary \\
 \qquad \Bwhen biscuit \in chosen 
\; \Bthen chosen := chosen - \{biscuit\}
    \Bend \\ 
 \quad \mbox{dispenseChoc} \defs   \Bstatus: \Bordinary\\
 \qquad \Bwhen choc \in chosen 
\; \Bthen  chosen := chosen - \{choc\}
    \Bend \\ 
\Bend
\end{array}$ }
\end{center}
\caption{$VM_1$}
\label{fig:VM_1}
\end{figure}

In Section~\ref{sec:strategies} we introduced a development strategy for a vending machine. Figures~\ref{fig:VM_1},~\ref{fig:VM_2},~\ref{fig:VM_3} and~\ref{fig:VM_4} illustrate a development chain from vending machine $VM_1$, $VM_2$, $VM_3$ to $VM_4$; there are no anticipated events in $VM_4$. Note the numbers of the vending machines start from one. We introduce $VM_0$ in Section~\ref{sec:extendingPreserving}. Thus $M_0$ in Theorem~\ref{thm:chain_ca} corresponds to $VM_1$ etc.

$VM_1$ is a simple machine that supports the selection and dispensing of chocolates and biscuits via four events: {\it selectBiscuit}, {\it selectChoc}, {\it dispenseBiscuit} and {\it dispenseChoc}. We abbreviate their names in the narrative to {\it sb}, {\it sc}, {\it db} and {\it dc} respectively. The first refinement step introduces $VM_2$ and the notion of paying and refunding.  The {\it pay} event in $VM_2$  is always enabled and allows positive credit to be input. The machine allows a biscuit to be chosen if it has not already been chosen and additionally provided a payment has been made; a chocolate selection is similar. Hence the guards of all four of the original events {\it sb}, {\it sc}, {\it db} and {\it db} are strengthened. The guard of the {\it refund} event means that credit cannot be refunded for selected items and cannot occur forever since it is convergent. Importantly, the {\it refundEnabled} flag is introduced so that it is only true after a dispense and prevents infinite loops of the {\it pay} followed by {\it  refund}.

\begin{figure}[ht!]
\begin{center} 
\framebox{
$\begin{array}{l}
\Bmachine VM_2 \\
\Bvariables credit, chosen, refundEnabled\\ 
\Binvariant credit \in \nat \land chosen \subseteq\{ choc, biscuit \} \land refundEnabled \in BOOL\\
\Bvariant if \; refundEnabled = FALSE \; then \; 0 \; else \; 1\\
\Bevents \\
\quad \mbox{init} \defs 
   credit := 0 || chosen := \{\} || refundEnabled := FALSE\\
 \quad \mbox{pay} \defs   \Bstatus: \Banticipated\\
 \qquad \Bany x \Bwhere x \in \nat_1 \\
\qquad \Bthen credit := credit + x  \Bend ||  refundEnabled := FALSE
    \Bend \\ 
 \quad \mbox{selectBiscuit} \defs  \Bstatus: \Bordinary\\
 \qquad \Bwhen credit > 0 \land biscuit \nin chosen \land  credit > card(chosen) \\
\qquad \Bthen chosen := chosen \union \{ biscuit \}  
    \Bend \\ 
 \quad \mbox{selectChoc} \defs \Bstatus: \Bordinary\\
 \qquad \Bwhen credit > 0 \land choc \nin chosen \land credit >  card(chosen) \\
\qquad \Bthen chosen := chosen \union \{ choc \} 
    \Bend \\ 
 \quad \mbox{dispenseBiscuit} \defs   \Bstatus: \Bordinary \\
 \qquad \Bwhen credit > 0 \land biscuit \in chosen \\
\qquad \Bthen credit := credit - 1 || chosen := chosen - \{biscuit\} || \\
\qquad \qquad \quad refundEnabled := TRUE 
    \Bend \\ 
 \quad \mbox{dispenseChoc} \defs \Bstatus: \Bordinary \\
 \qquad \Bwhen credit > 0 \land choc \in chosen \\
\qquad \Bthen  credit  := credit - 1 || chosen := chosen - \{choc\} || \\
\qquad \qquad  \quad refundEnabled := TRUE 
    \Bend \\ 
    \quad \mbox{refund} \defs   \Bstatus: \Bconvergent \\
 \qquad \Bwhen  credit > card(chosen) \land refundEnabled := TRUE\\
\qquad \Bthen  credit  := card(chosen) || refundEnabled := FALSE
  \Bend \\ 
\Bend
\end{array}$ }
\end{center}
\caption{$VM_2$}
\label{fig:VM_2}
\end{figure} 

$VM_3$ introduces the notion of stocked items and a new {\it refill} event. We could have chosen many different guards for the {\it refill} event. For example, we could have labelled it {\it anticipated} with a guard of {\it true}. Instead we have made an underspecification where the stock can be restocked when there may be no biscuits or no chocolates, and established convergence. Again the  guard of the four original events have been strengthened so that they are only enabled when the appropriate stocked item is in stock. But now {\it db} and {\it dc} also capture the non-deterministic notion of running out or not of their respective items. The guard of {\it refund} remains unchanged. The guard of {\it pay} has been strengthened so that it is only enabled when there is stock but this is not strong enough to prevent it happening infinitely often, hence it remains anticipated in $VM_3$.

The final machine, $VM_4$, is a straightforward data refinement which introduces the capacity of the machine. Apart from highlighting the refinement relationship between $stocked$ and $chocStock$ and $biscuitStock$  note the strengthening of the guard of {\it refill} so that  vending machine should only be refilled when there is no stock. Also the guard of {\it pay} is strengthened so that it becomes convergent.

\begin{figure}[ht!]
\begin{center} 
\framebox{
$\begin{array}{l}
\Bmachine VM_3 \\
\Bvariables credit, chosen, refundEnabled, stocked  \\ 
\Binvariant credit \in \nat \land chosen \subseteq \{choc, biscuit\} \land stocked \subseteq \{choc, biscuit\}\\
\qquad \qquad (choc \in chosen \implies choc \in stocked) \land (biscuit \in chosen \implies biscuit \in stocked)\\
\Bvariant  card\{choc,biscuit\}-stocked\\
\Bevents \\
\quad \mbox{init} \defs 
 \ldots || stocked := \{choc, biscuit\}\\
 \quad \mbox{pay} \defs   \Bstatus: \Banticipated\\
 \qquad \Bany x \Bwhere x \in \nat_1 \land stocked \neq \emptyset \\
\qquad \Bthen credit := credit + x  \Bend ||   refundEnabled := FALSE
    \Bend \\ 
 \quad \mbox{selectBiscuit} \defs \Bstatus: \Bordinary  \\
 \qquad \Bwhen \ldots \land biscuit \in stocked\\
\qquad \Bthen chosen := chosen \cup \{biscuit\}  \Bend \\ 
\quad \mbox{selectChoc} \defs  \Bstatus: \Bordinary \\
 \qquad \Bwhen \ldots \land choc \in stocked \\
\qquad \Bthen  chosen  := chosen \cup \{ choc \} \Bend \\ 
\quad \mbox{dispenseBiscuit} \defs  \Bstatus: \Bordinary \\
 \qquad \Bwhen credit > 0 \land biscuit \in chosen \land biscuit \in stocked\\
\qquad  \Bthen \ldots || \Bany \thinspace x \thinspace \Bwhere \thinspace x \subseteq \{biscuit\} 
\Bthen stocked := stocked \setminus x \Bend
  \Bend \\  
\quad \mbox{dispenseChoc} \defs  \Bstatus: \Bordinary \\
 \qquad \Bwhen credit > 0  \land choc \in chosen \land choc \in stocked\\
\qquad  \Bthen \ldots || \Bany \thinspace x \thinspace \Bwhere \thinspace x \subseteq \{choc\}  \Bthen stocked := stocked \setminus x \Bend

    \Bend \\ 
  \quad \mbox{refund} \defs  \Bstatus: \Bordinary  \ldots\\
         \quad \mbox{refill} \defs \\  
\qquad  \Bstatus: \Bconvergent \\
 \qquad \Bwhen choc \notin stocked  \lor biscuit \notin stocked\\
\qquad \Bthen  stocked  := \{choc, biscuit\}
     \Bend \\
     
\Bend
\end{array}$ }
\end{center}
\caption{$VM_3$}
\label{fig:VM_3}
\end{figure}

\begin{figure}[t!]
\begin{center} 
\framebox{
$\begin{array}{l}
\Bmachine VM_4 \\
\Bconstants capacity\\
\Bproperties capacity >0  \\
\Bvariables credit, chosen, refundEnabled, chocStock, biscuitStock \\ 
\Binvariant credit \leq capacity \land chosen \subseteq \{choc, biscuit \} \land  \\
\qquad  refundEnabled \in BOOL \land chocStock \leq capacity \land biscuitStock \leq capacity \land \\
\qquad  (choc \notin stocked \implies chocStock = 0 ) \land 
  (choc \in stocked \implies chocStock \geq 0) \land \\
\qquad ( biscuit \notin stocked \implies  biscuitStock = 0) \land 
  (biscuit \in stocked \implies biscuitStock \geq 0) \\
\Bvariant  max\{(chocStock + biscuitStock) - credit,0\}\\
\Bevents \\
\quad \mbox{init} \defs 
\ldots || chocStock := capacity || biscuitStock := capacity \\
 \quad \mbox{pay} \defs  \Bstatus: \Bconvergent \\
 \qquad \Bany x \Bwhere x \in \nat_1 \land (chocStock + biscuitStock) > credit \\
\qquad \Bthen credit := credit + x  \Bend ||  refundEnabled := FALSE
\Bend \\
\quad \mbox{selectChoc} \defs  \Bstatus: \Bordinary \\
 \qquad \Bwhen \ldots  \land chocStock > 0 \\
\qquad \Bthen  chosen  := chosen \cup \{ choc \} \Bend \\ 
 \quad \mbox{selectBiscuit} \defs   \Bstatus: \Bordinary \\
 \qquad \Bwhen \ldots \land biscuitStock > 0 \\
\qquad \Bthen chosen := chosen \cup \{biscuit\}  \Bend \\ 
 \quad \mbox{dispenseBiscuit} \defs  \Bstatus: \Bordinary\\
 \qquad \Bwhen credit > 0 \land biscuit \in chosen \land  biscuitStock > 0\\
\qquad \Bthen \ldots || chocStock := chocStock - 1

  \Bend \\ 
 \quad \mbox{dispenseChoc} \defs   \Bstatus: \Bordinary \\
 \qquad \Bwhen credit > 0 \land choc \in chosen \land  chockStock > 0\\
\qquad \Bthen \ldots || chocStock := chocStock - 1
  
   \Bend \\ 
  \quad \mbox{refund} \defs \Bstatus: \Bordinary \ldots\\
       \quad \mbox{refill} \defs   \Bstatus: \Bordinary \\
 \qquad \Bwhen chocStock = 0 \land biscuitStock = 0 \\
\qquad \Bthen  chocStock := capacity || biscuitStock := capacity
     \Bend \\
\Bend
\end{array}$ }
\end{center}
\caption{$VM_4$}
\label{fig:VM_4}
\end{figure} 


\section{LTL notation}
\label{sec:ltl}

In this paper we use the grammar for the LTL operators presented by Plagge and Leuschel~\cite{prob_ltl}:
\begin{eqnarray*}
\phi & ::= & true \mid [x]\mid \neg \phi \mid \phi_1 \vee \phi_2 \mid \phi_1 \negthinspace \;\ U \; \phi_2
\end{eqnarray*}
A machine $M$ satifies $\phi$, denoted $M \models \phi$, if all traces of $M$ satisfy $\phi$.  The definition for $u$ to satisfy $\phi$ is defined by induction over $\phi$ as follows:
\[
\begin{array}{lcl}
u \models true \\
u \models [x]  & {}\Leftrightarrow{} &  u = \langle x \rangle \cat u^{1} \\
u \models \neg \phi  & {}\Leftrightarrow{} &  \mbox{it is not the case that $u \models \phi$} \\
u \models \phi_1 \lor \phi_2 & {}\Leftrightarrow{} &  u \models \phi_1 \;\mbox{or}\; u \models \phi_2 \\
u \models \phi_1 U \phi_2  & {}\Leftrightarrow{} &  \exists k \geq 0 .  \forall i < k . u^{i} \models \phi_1 \;\mbox{and}\; u^{k} \models \phi_2
\end{array} 
\]
where $u^{n}$ is $u$ with the first $n$ elements removed, i.e., $u = \trace{x_0,\ldots,x_{n-1}} \cat u^{n}$.

From these operators Plagge and Leuschel derived several additional operators, including: conjunction ($\phi_1 \land \phi_2$), finally (or eventually) ($F \phi$), and globally (or always) ($G \phi$), in the usual way; we also use these operators, and for explicitness we also provide direct definitions for them:
\[
\begin{array}{lcl}
u \models \phi_1 \land \phi_2 & {}\Leftrightarrow{} & u \models \phi_1 \;\mbox{and}\; u \models \phi_2 \\
u \models  F \phi  & \Leftrightarrow & \exists i \geq 0 .  u^{i} \models \phi \\
u \models  G \phi & \Leftrightarrow & \forall i \geq 0 .  u^{i} \models \phi
\end{array} 
\]
We omit atomic propositions on states since our traces are only dealing with events and not paths of states and transitions. We also omit the next operator, see Section~\ref{sec:related_work}. In this paper our running example uses globally, finally, or and implies. 

For example, the informal specification for the $Lift$ given in Section~\ref{sec:intro}, that whenever $top$ happens then eventually $ground$ will happen, could be written as
\[ G([top] \implies F [ground]) \]
From our running $VM$ example, the predicate $G F [selectBiscuit]$ expresses that $selectBiscuit$ occurs infinitely often: at any point there is always some occurrence of $selectBiscuit$ at some point in the future.  We use this construction in the VM properties introduced in Section~\ref{sec:preserving}.  For example, we have $\phi_2$ given as
\[ \phi_2 \;\;=\;\; (\neg G F [selectBiscuit]) \implies G([selectChoc] \implies F[dispenseChoc]) \]
This states that provided $selectBiscuit$ only occurs finitely often (i.e. eventually stops), then whenever $selectChoc$ occurs then $dispenseChoc$ will eventually occur.
 
It will also be useful to identify the events mentioned explicitly in an LTL formula $\phi$.  This set is called the alphabet of $\phi$.  This is written $\alpha(\phi)$, similar to the use of $\alpha M$ for the alphabet of machine $M$.  For LTL formulae it is defined inductively as follows:
\begin{definition}
\label{defn:alpha}
\begin{eqnarray*}
\alpha(true) &  = & \{ \} \\
\alpha([x]) & = & \{ x \} \\
\alpha(\neg \phi) & = & \alpha(\phi) \\
\alpha(\phi_1 \lor \phi_2) & = & \alpha(\phi_1) \union \alpha(\phi_2) \\
\alpha(\phi_1 \land \phi_2) & = & \alpha(\phi_1) \union \alpha(\phi_2) \\
\alpha(\phi_1 \; U \; \phi_2) & = & \alpha (\phi_1) \; \union \; \alpha(\phi_2) \\
\alpha(F \phi) & = & \alpha(\phi)\\
\alpha(G \phi) & = & \alpha(\phi)
\end{eqnarray*}
\end{definition}
For example, we have $\alpha(\phi_2) = \{ selectBiscuit, selectChoc, dispenseChoc \}$ for $\phi_2$ above.

\section{Preserving LTL properties}
\label{sec:preserving}

In this section we provide results to demonstrate when properties are preserved by refinement chains. Firstly, we consider chains which do not contain any renaming/splitting of events in a machine. Hence, each function mapping $f_i$ for $M_i \ldots M_n$ is the identity.
The first result is a general result identifying a particular temporal property that will always hold for all refinement chains which abide by the rules of the strategy presented in Section~\ref{sec:strategies}. The second result given in Lemma~\ref{lemma:chain_ltlpreserve} concerns the preservation of temporal properties that would be proposed by a specifier. We have already observed from the vending machine example that new events can be introduced during a refinement, e.g., {\it pay}, {\it refill},  etc.. We aim for such properties to hold even though new anticipated and convergent events are being introduced. 


Lemma~\ref{lemma:chain_gfoprop} states that $M_n$ at the end of the refinement chain will always eventually perform one of the  events of the initial machine $M_0$. In other words, $M_n$ will perform infinitely many of the initial events.  This means that the events introduced along the refinement chain cannot occur forever at the expense of the original events.  In our example, $\alpha M_0 = O_0$.

\begin{lemma}
\label{lemma:chain_gfoprop}
If $M_0 \Weventref M_1 \Weventref \ldots \Weventref M_n$ and $M_n$ is deadlock free and $M_n$ does not contain any anticipated events then $M_n \models GF(\underset{e \in \alpha M_0}{\bigvee} \; [e] )$\\
\end{lemma}
%

Next we provide a definition which is used in Lemma~\ref{lemma:chain_ltlpreserve} below and it enables us to gain insights into the kinds of temporal properties that are appropriate to be proposed and have the potential of being preserved through a refinement chain. Definition~\ref{defn:bDependent} describes a maximal execution satisfying a property $\phi$. The execution may include some events which do not have an impact on whether the property holds or not therefore we can restrict the maximal execution to include only those events that impact on the property. 
\begin{definition}
\label{defn:bDependent}
Let $\beta$ be a set of events. Then 
$\phi$ is $\beta$-dependent  if  $\alpha(\phi) \subseteq \beta$ and \\$u \models \phi \Leftrightarrow (u \project \beta) \models \phi$.
\end{definition}

An example of a $\beta$-dependent property is $GF(pay)$ where $\beta = \{pay\}$. If $u \models GF(pay)$ then $u \project{\it pay} \models GF(pay)$, and vice versa.
Conversely, $\neg G(pay)$ is not $\{pay\}$-dependent. For example, if $u=\trace{pay,refill,pay,pay,\ldots}$ then $u \models \neg G(pay)$ but $u \project \{pay\} \not \models \neg G(pay)$.


As another example, define $\beta = \{sb,sc,db,dc\}$.  Then $G(sb \vee sc \vee db \vee dc)$ is not $\beta$-dependent. This is exemplified by any trace $u$ which contains events other than those in $\beta$.  In this case $u \project \{sb,sc,db,sc\} \models G(sb \vee sc \vee db \vee dc)$  but $u \not \models  G(sb \vee sc \vee db \vee dc)$.   $VM_4$ exhibits such traces.  Observe that this property holds for $VM_1$ but not for $VM_4$: it is not preserved by refinement.   Since it is not $\beta$-dependent Lemma~\ref{lemma:chain_ltlpreserve} below is not applicable for this property.

Our main result for this section identifies conditions under which an LTL property $\phi$ will be preserved in a refinement chain.  The conditions are as follows:
\begin{itemize}
\item by the end of the refinement chain there should be no outstanding anticipated events (and so all newly introduced events have been shown to be convergent), as given by restriction~6 of the Development Strategy of Section~\ref{subsec:dev};
\item the final machine in the refinement chain must be deadlock-free; and
\item all of the events that have an effect on whether or not $\phi$ is true are already present in $M_i$ ($\phi$ is $\beta$-dependent for some $\beta \subseteq \alpha M_i$).
\end{itemize}
These conditions are enough to ensure that $\phi$ is preserved through refinement chains.  This means that $M_i$ can be checked for $\phi$, and we can be sure that the resulting system $M_n$ will also satisfy it.

The lemma is formally expressed as follows:
\begin{lemma}
\label{lemma:chain_ltlpreserve}
If $M_i \models \phi$ and $M_i \Weventref \ldots \Weventref M_n$ and $0 \leq i < n$ and $M_n$ is deadlock free and $M_n$ does not contain any anticipated events and $\phi$ is $\beta$-dependent and $\beta \subseteq \alpha M_i$ then $M_n \models \phi$.
\end{lemma}


\subsection{Preserving Vending Machine properties}
\label{sec:example_preservation}

We consider the application of the above Lemmas to our running example on the refinement chain 
\[ VM_1 \Weventref VM_2 \Weventref VM_3 \Weventref VM_4 \]
In this case we obtain immediately from Lemma~\ref{lemma:chain_gfoprop} that 
\[ VM_4 \models G F ([selectBiscuit] \lor [selectChoc] \lor \\ \hspace*{0.765in} [dispenseBiscuit] \lor [dispenseChoc]) \]
Any execution of $VM_4$ will involve infinitely many occurrences of some of these events.  The newly introduced events $pay$, $refund$, $refill$ cannot be performed forever without the occurrence of the original events.

We consider some further properties to illustrate the applicability of Lemma~\ref{lemma:chain_ltlpreserve}.  Taking $VM_1$ to be the first machine in the refinement chain, we can consider the following temporal properties $\phi$ for $VM_1$:
\begin{eqnarray*}
\phi_{1} &=& G([selectChoc] \vee [selectBiscuit] \implies F([dispenseChoc] \vee [dispenseBiscuit]))\\
\phi_{2} &=& (\neg GF [selectBiscuit]) \implies G([selectChoc] \implies F [dispenseChoc])\\
\phi_{3} &=& (\neg GF [selectChoc]) \implies G([selectBiscuit] \implies F [dispenseBiscuit])\\
\phi_{4} &=&  G([selectChoc] \implies  F [dispenseChoc])\\
\phi_{5} &=& G([selectBiscuit] \implies F [dispenseBiscuit])
\end{eqnarray*}
We note that each of the properties are $\beta$-dependent. Next we consider whether $VM_1 \models \phi_i$ for each $i \in {1..5}$. Note that in fact $VM_1 \not \models \phi_4$ and $VM_1 \not \models \phi_5$ since there is a trace for which the properties fail, e.g., in the case of $\phi_4$  the $\trace{sc,sb,db,sb,db,\ldots}$ we could have an infinite loop of $sb,db$ events and never reach a $dc$ event.  Thus Lemma~\ref{lemma:chain_ltlpreserve} is not applicable to these properties.

The properties $\phi_2$ and $\phi_3$ are the strongest; $\phi_2$ states that if you do not always have an $sb$ then you will be able to choose a chocolate and for it to be dispensed, and the dual applies in $\phi_3$. Once we have also established the refinement chain $VM_1 \Weventref VM_2 \Weventref VM_3 \Weventref VM_4$,  and that $VM_4$ is deadlock free we can deduce using Lemma~\ref{lemma:chain_ltlpreserve} that $VM_4 \models \phi_i$ for all $i \in 1..3$. Observe however that Lemma~\ref{lemma:chain_ltlpreserve} does not establish that  $\phi_i$ holds in all refinement machines, only those with no anticipated events. For example, $VM_2$ and $VM_3$ do not satisfy $\phi_1$, $\phi_2$ nor $\phi_3$ since {\it pay} is anticipated and can be executed infinitely often. 

Since $VM_2$ introduced the event $pay$ we can also introduce new temporal properties that are required to hold from $VM_2$ onwards.  In other words, we apply Lemma~\ref{lemma:chain_ltlpreserve} on the chain $VM_2 \Weventref VM_3 \Weventref VM_4$.  The properties to consider are:
\begin{eqnarray*}
\phi_{6} &=& G([pay] \implies F ([dispenseBiscuit] \vee [dispenseChoc]))\\
\phi_{7} &=& GF[pay]
\end{eqnarray*}
The infinite behaviour of $pay$ means that $\phi_6$ is not satisfied in $VM_2$. However, $VM_2 \models \phi_7$ thus we can again apply Lemma~\ref{lemma:chain_ltlpreserve}, and obtain that $VM_4 \models \phi_7$ since $\phi_7$ is $\beta$-dependent. This exemplifies that new temporal properties can be added to the refinement verification chain. 

We note that in fact $VM_4 \models \phi_6$. Thus $\phi_6$ and $\phi_7$ together imply that $GF([dispenseBiscuit] \vee [dispenseChoc]))$ holds for $VM_4$. 



\section{Extending preserving LTL properties to handle splitting events}
\label{sec:extendingPreserving}
In this section we generalise the results of Section~\ref{sec:preserving} in order to deal with splitting events in Event-B, which occurs when abstract events are refined by several events in the concrete machine, corresponding to a set of alternatives.   Consider as a motivating example $VM_0$ in Figure~\ref{fig:VM_0}.  This is refined by $VM_1$, with linking invariant $item = card(chosen)$,  {\em selectItem} refined by both {\em selectBiscuit} and {\em selectChoc}, and {\em dispenseItem} refined by both {\em dispenseBiscuit} and {\em dispenseChoc}.  Splitting events also involves their renaming to allow for several concrete events to map to the same abstract one.  A refinement step will therefore be associated with a renaming function $h$ from concrete events to the abstract events that they refine.  In the general case $h$ will be many-to-one, since many concrete events may map to a single abstract event; and it will also be partial, since new events in the concrete machine will not map to any abstract event.  

In general, each step in a refinement chain $M_0\Weventref M_1 \Weventref \ldots \Weventref M_n$ will have an event renaming function $h_i$ corresponding to the renaming and splitting step from $M_{i}$ to $M_{i-1}$.  We define $g_{i,n}$ to be the composition of these renaming function from $h_n$ down to $h_i$.  Observe that $g_{i,n}$ will be undefined on any event that does not map to $M_{i-1}$, in other words any event that corresponds to an event introduced at some point in the refinement chain.  For example, for the chain $VM_0 \Weventref VM_1 \Weventref \ldots \Weventref VM_4$, we obtain that $g_{1,4}(selectBiscuit) = g_{1,4}(selectChoc) = selectItem$, and $g_{1,4}(dispenseBiscuit) = g_{1,4}(dispenseChoc) = dispenseItem$, and $g_{1,4}$ is not defined on the remaining events of $VM_4$.


Lemma~\ref{lemma:chain_gfoprop} generalises to state that the final machine in the refinement chain must always eventually perform some event relating to an event in the initial machine.
\begin{lemma}
\label{lemma:chain_gfoprop_extended}
If $M_0 \Weventref M_1 \Weventref \ldots \Weventref M_n$ and $M_n$ is deadlock free and $M_n$ does not contain any anticipated events then $M_n \models GF(\underset{e \in g_{1,n}^{-1}(\alpha M_0)}{\bigvee} \; e )$\\
\end{lemma}
Observe that if there is no renaming or splitting, then $g_{1,n}$ is the identity function on the events in $\alpha M_0$, yielding Lemma~\ref{lemma:chain_gfoprop}.

We are interested in how the LTL properties of an abstract machine becomes transformed through a refinement step such as $VM_0$ to $VM_1$.  For example, the property $G F [selectItem]$ for $VM_0$ states that from any stage that is reached, $selectItem$ will eventually occur.  This will translate to the property $GF([selectBiscuit] \vee [selectChoc])$ for $VM_1$.  We now consider how LTL properties translate through a renaming function $h$.

\begin{figure}[t!]
\begin{center} 
\framebox{
$\begin{array}{l}
\Bmachine VM_0 \\
\Bvariables item\\ 
\Binvariant  item \in \nat \\
\Bevents \\
\quad \mbox{init} \defs  item :=  0 \\
 \quad \mbox{selectItem} \defs \\  
 \qquad  \Bstatus: \Bordinary \\
 \qquad \Bwhen item \leq 2 
 \; \Bthen item :=item +1  \Bend \\ 
 \quad \mbox{dispenseItem} \defs \\  
\qquad  \Bstatus: \Bordinary \\
 \qquad \Bwhen item > 0  
\; \Bthen item := item -1  \Bend \\ 
\Bend
\end{array}$ 
}
\end{center}
\caption{$VM_0$}
\label{fig:VM_0}
\end{figure} 

For a given event renaming function $h$, we define $trans_h$ as the translation that maps LTL formulae by mapping abstract events to the disjunction of their corresponding concrete events, as follows:
\begin{definition}
\label{defn:trans}
\begin{eqnarray*}
trans_h(true) &  = & true \\
trans_h([x]) & = & \underset{y  \mid h(y) = x}{\bigvee} [y] \\
trans_h(\neg \phi) & = & \neg trans_h(\phi) \\
trans_h(\phi_1 \lor \phi_2) & = & trans_h(\phi_1) \lor trans_h(\phi_2) \\
trans_h(\phi_1 \land \phi_2) & = & trans_h(\phi_1) \land trans_h(\phi_2) \\
trans_h(\phi_1 \; U \; \phi_2) & = & trans_h (\phi_1) \; U \; trans_h(\phi_2) \\
trans_h(G \phi) & = & G \; trans_h(\phi) \\
trans_h(F \phi) & = & F \; trans_h(\phi)
\end{eqnarray*}
\end{definition}
For example
\begin{eqnarray*}
\lefteqn{trans_h(G ([selectItem] \implies F [dispenseItem]))} \\ & = & 
G (([selectBiscuit] \lor [selectChoc]) \implies F ([dispenseBiscuit] \lor [dispenseChoc])) 
\end{eqnarray*}
%

Lemma~\ref{lemma:chain_ltlpreserve} generalises to Lemma~\ref{lemma:chain_ltlpreserve_extended} below, to state that
 LTL properties are carried along the refinement chain by translating them.  In particular, if a property $\phi$ is established for $M_{i-1}$, then $trans_{g_{i,n}}(\phi)$ will hold for $M_n$:
\begin{lemma}
\label{lemma:chain_ltlpreserve_extended}
If $M_{i-1} \models \phi$ and $M_{i-1} \Weventref \ldots \Weventref M_n$ and $0 \leq i-1 < n$, $M_n$ is deadlock free and $M_n$ does not contain any anticipated events and $\phi$ is $\beta$-dependent and $\beta \subseteq \alpha M_{i-1}$ then $M_n \models trans_{g_{i,n}}(\phi)$
\end{lemma}
For example, from the result for $VM_0$ that whenever {\em selectItem} occurs then {\em dispenseItem} will eventually occur,
\begin{eqnarray*}
VM_0 & \models & G ([selectItem] \implies F [dispenseItem]))
\end{eqnarray*}
we obtain from Lemma~\ref{lemma:chain_ltlpreserve_extended} that 
\begin{eqnarray*}
VM_4 & \models &
G (\; ([selectBiscuit] \lor [selectChoc]) \\
&& \hspace*{0.4in} {} \implies F ([dispenseBiscuit] \lor [dispenseChoc]) \; ) 
\end{eqnarray*}
This states that whenever {\em selectBiscuit} or {\em selectChoc} occur, then {\em dispenseBiscuit} or {\em dispenseChoc} will eventually occur.

\section{Discussion and related work}
\label{sec:related_work}
%
%

One of the few papers to discuss LTL preservation in Event-B refinement is Groslambert~\cite{B-LTL-Groslambert}.  The LTL properties were defined in terms of predicates on system state rather than our paper's formulation in terms of the occurrence of events. His paper focused only on the introduction of new convergent events. It did not include a treatment of anticipated events but this is unsurprising since the paper was published before their inclusion in Event-B. Our results are more general in two ways. Firstly, the results support the treatment of anticipated events. Secondly, we allow more flexibility in the development methodology. A condition of Groslambert's results was that all the machines in the refinement chain needed to be deadlock free. The two main lemmas in our paper:  Lemmas~\ref{lemma:chain_ltlpreserve} and~\ref{lemma:chain_ltlpreserve_extended} do not require each machine in a refinement chain to be deadlock free, only the final machine.  It is irrelevant if intermediate $M_i$s deadlock as long as the deadlock is eventually refined away.


Groslambert deals with new events via stuttering and leaves them as visible events in a trace. This is why the LTL operators used by the author do not include the next operator ($X$). As new events may happen this may violate the $X$ property to be checked. Plagge and Leuschel in~\cite{prob_ltl} permit the use of the $X$ operator since they treat the inclusion of new events as internal events which are not visible. Since we deal with new events as visible events we also lose the ability to reason about a temporal property using the typical $X$ operator. Our reasoning is simpler than both Groslambert and Plagge and Leuschel since we only focus on events but this means we cannot have atomic propositions in our LTL, whereas they can.

The notion of verification of temporal properties of both classical and Event-B systems using proof obligations has been considered in many research papers. Abrial and Musat in an early paper,~\cite{DBLP:conf/b/AbrialM98}, introduced proof obligations to deal with dynamic constraints in classical B. In a more recent paper~\cite{DBLP:conf/icfem/HoangA11} Hoang and Abrial have also proposed new proof obligations for dealing with liveness properties in Event-B. They focus on three classes of properties: existence, progress and persistence, with a view to implementing them in Rodin.  Bicarregui {\it et al.} in~\cite{Bicarregui-B} introduced a temporal concept into events using the guard in the {\it when} clause and the additional labels of {\it within} and {\it next} so that the enabling conditions are captured clearly and separately. However, these concepts are not aligned with the standard Event-B labelling.


The interest of LTL preservation through refinement is wider than simply Event-B. Derrick and Smith~\cite{DBLP:journals/fac/DerrickS12} discuss the preservation of LTL properties in the context of Z refinement but the authors extend their results to other logics such as CTL and the $\mu$ calculus. They focus on discussing the restrictions that are needed on temporal-logic properties and retrieve relations to enable the model checking of such properties. Their refinements are restricted to data refinement and do not permit the introduction of new events in the refinement steps. Our paper does permit new events to be introduced during refinement steps; the contribution is in identifying conditions for LTL properties to hold even in the context of such new events.

\section{Conclusions and future work}
\label{sec:future_work}
The paper has provided foundational results that justify when temporal properties hold at the end of an Event-B refinement chain for developments which contain anticipated, convergent and ordinary events, which goes beyond that presented in~\cite{B-LTL-Groslambert}.  The paper has also provided restrictions on the temporal properties in terms of being $\beta$-dependent which help to determine when a temporal property of interest should be introduced into the development chain.

We could extend the results to deal with merging events. The inclusion of the $X$ LTL operator and availability will require use to look at execution paths which include state transitions ($\pi$ paths). The inclusion of availability will enable us to address more advanced and useful notions of fairness in the context of temporal properties. Our notion of weak fairness will be akin to that described in Barradas and Bert  in~\cite{didier-fairnessB}. It will draw on work by Williams {\it et al.}~\cite{DBLP:conf/ictac/WilliamsRF12}. We could also consider the impact on temporal property preservation in refinement chains which do  not achieve convergence of all its new events by the end.

In ongoing work we are looking at event liveness via the proof obligation for strong deadlock freedom {\bf \tt S\_NDF}. We have defined new labelling of events to so that liveness proofs are on particular events. This is analagous to proving {\bf \tt WFD\_REF} for events that are labelled anticipated or convergent. We have recently defined the semantics of Event-B in terms of stable failures and detailed its relationship with {\bf \tt S\_NDF}. We are currently combining these results with our work in~\cite{DBLP:journals/fac/SchneiderTW14} in order to provide a cohesive process algebra underpinning for Event-B.



\ \\ \noindent{\bf Acknowledgments.} 
Thanks to Thai Son Hoang and Thierry Lecomte for
 discussions about Event-B development strategies and the challenges of discharing liveness proofs. Thanks to Steve Wesemeyer for discussions on the example. Thanks to the reviewers for their constructive comments that helped to improve the paper.

\bibliographystyle{abbrv}
\bibliography{ifm2014_bib}

\begin{thebibliography}{10}

\bibitem{Abrial09}
J.-R. Abrial.
\newblock {\em Modeling in {E}vent-{B}: System and Software Engineering}.
\newblock Cambridge University Press, 2010.

\bibitem{DBLP:journals/sttt/AbrialBHHMV10}
J.-R. Abrial, M.~J. Butler, S.~Hallerstede, T.~S. Hoang, F.~Mehta, and
  L.~Voisin.
\newblock Rodin: an open toolset for modelling and reasoning in {E}vent-{B}.
\newblock {\em STTT}, 12(6):447--466, 2010.

\bibitem{DBLP:conf/b/AbrialM98}
J.-R. Abrial and L.~Mussat.
\newblock Introducing dynamic constraints in {B}.
\newblock In {\em B}, volume 1393 of {\em LNCS}, pages 83--128. Springer, 1998.

\bibitem{didier-fairnessB}
H.~Barradas and D.~Bert.
\newblock Specification and proof of liveness properties under fairness
  assumptions in {B} event systems.
\newblock In {\em Integrated Formal Methods}, volume 2335 of {\em LNCS}, pages
  360--379. Springer, 2002.

\bibitem{Bicarregui-B}
J.~Bicarregui, A.~Arenas, B.~Aziz, P.~Massonet, and C.~Ponsard.
\newblock Towards modelling obligations in {E}vent-{B}.
\newblock In {\em Abstract State Machines, B and Z}, volume 5238 of {\em LNCS},
  pages 181--194. Springer, 2008.

\bibitem{butlerphd}
M.~J. Butler.
\newblock {\em A {CSP} approach to Action Systems}.
\newblock {DP}hil thesis, Oxford U., 1992.

\bibitem{DBLP:journals/fac/DerrickS12}
J.~Derrick and G.~Smith.
\newblock Temporal-logic property preservation under {Z} refinement.
\newblock {\em Formal Asp. Comput.}, 24(3):393--416, 2012.

\bibitem{B-LTL-Groslambert}
J.~Groslambert.
\newblock Verification of {LTL} on {B} {E}vent {S}ystems.
\newblock In {\em B 2007: Formal Specification and Development in B}, volume
  4355 of {\em LNCS}, pages 109--124. Springer, 2006.

\bibitem{Hallerstede2013272}
S.~Hallerstede, M.~Leuschel, and D.~Plagge.
\newblock Validation of formal models by refinement animation.
\newblock {\em Science of Computer Programming}, 78(3):272 -- 292, 2013.

\bibitem{DBLP:conf/icfem/HoangA11}
T.~S. Hoang and J.-R. Abrial.
\newblock Reasoning about liveness properties in {E}vent-{B}.
\newblock In {\em ICFEM}, volume 6991 of {\em LNCS}, pages 456--471. Springer,
  2011.

\bibitem{DBLP:journals/sttt/LeuschelB08}
M.~Leuschel and M.~J. Butler.
\newblock Pro{B}: an automated analysis toolset for the {B} method.
\newblock {\em STTT}, 10(2):185--203, 2008.

\bibitem{DBLP:conf/fm/LeuschelFFP09}
M.~Leuschel, J.~Falampin, F.~Fritz, and D.~Plagge.
\newblock Automated property verification for large scale {B} models.
\newblock In {\em FM}, volume 5850 of {\em LNCS}, pages 708--723. Springer,
  2009.

\bibitem{morgan}
C.~Morgan.
\newblock Of wp and {CSP}.
\newblock {\em Beauty is our business: a birthday salute to E. W. Dijkstra},
  pages 319--326, 1990.

\bibitem{prob_ltl}
D.~Plagge and M.~Leuschel.
\newblock Seven at one stroke: {LTL} model checking for high-level
  specifications in {B}, {Z}, {CSP}, and more.
\newblock {\em STTT}, 12(1):9--21, 2010.

\bibitem{DBLP:journals/fac/SchneiderTW14}
S.~Schneider, H.~Treharne, and H.~Wehrheim.
\newblock The behavioural semantics of {E}vent-{B} refinement.
\newblock {\em Formal Asp. Comput.}, 26(2):251--280, 2014.

\bibitem{DBLP:conf/ictac/WilliamsRF12}
D.~M. Williams, J.~de~Ruiter, and W.~Fokkink.
\newblock Model checking under fairness in {P}ro{B} and its application to fair
  exchange protocols.
\newblock In {\em ICTAC}, volume 7521 of {\em LNCS}, pages 168--182. Springer,
  2012.

\end{thebibliography}


\appendix
\section{Proofs of Lemmas}
\label{sec:proofs}


In this appendix we provide the proofs for the main lemmas: Lemma~\ref{lemma:chain_gfoprop}, ~\ref{lemma:chain_ltlpreserve} and Lemma~\ref{lemma:chain_ltlpreserve_extended}. We also provide additional supporting definitions and lemmas.

\noindent{\bf Proof of Lemma~\ref{lemma:chain_gfoprop}}
Since $M_n$ does not deadlock consider an infinite trace $u$ of $M_n$. Let $u = u \project (C_1 \cup \ldots \cup C_n \cup O_0)$ since $M_n$ does not have any anticipated events.  We aim to prove that $u \project O_0$ is infinite. If $u \project C_1 \cup \ldots \cup C_n$ is finite then $u \project O_0$ is infinite. If $u \project C_1 \cup \ldots \cup C_n$ is infinite then $u \project O_0$ is infinite by  Theorem~\ref{thm:chain_ca} where $g_{1,n}^{-1}$ is simply the identity function.
\hfill $\Box$ 
\\\

\noindent{\bf Proof of Lemma~\ref{lemma:chain_ltlpreserve}} 
Let  $u$ be a trace of $M_n$ and since $M_n$ is deadlock free $u$ will be infinite. Since $M_i \Weventref \ldots \Weventref M_n$  then we deduce that $u \project \alpha(M_i)$ is a trace of $M_i$. Therefore $u \project \alpha(M_i) \models \phi$. Therefore, $u \project \alpha(M_i) \project \beta \models \phi$, i.e., $u \project \beta \models \phi$, since $\beta \subseteq \alpha(M_i)$. By Definition~\ref{defn:bDependent} it follows that $u \models \phi$ as required.
\hfill $\Box$
 \\\

The following definition and lemmas regarding infinite traces and how they relate to temporal properties in the context of translation mappings: Definition~\ref{defn:htot}, Lemma~\ref{lem:h1}, Lemma~\ref{lemma:pathModelsPhi_extended} and Lemma~\ref{defn:bDependent_extended}, are required in order to prove Lemma~\ref{lemma:chain_ltlpreserve_extended} presented in Section~\ref{sec:extendingPreserving} and proved below.

Given a partial surjective function $h : B \psurj A$, we define the {\em completion} of $h$ as follows:
\begin{definition}
\label{defn:htot}
Given a partial surjection $h : B \psurj A$, its completion $h_{tot} : B \surj (A \union (B - dom(h)))$ is the function $h$ augmented with the identity on all events not in the domain of $h$.  $h_{tot}$ is defined as $h_{tot} = h \union id(B - dom(h))$.
\end{definition}

The $f_i$ ,$1 \leq i \leq n$, for each machine $M_i \ldots M_n$ are examples of $h$. $f_{2,tot}$ for $VM_2$ is $\{(pay,pay),(refund,refund),(sb,selectItem),(sc,selectItem), \ldots \}$.  
The \; function $h$ lifts to $B^\omega \psurj A^\omega$ by applying $h$ pointwise to each event.  $h_{tot}$ similarly lifts to $B^\omega \surj (A \union (B - dom(h)))^\omega$.

\begin{lemma} \label{lem:h1}
If $\alpha(\phi) \subseteq ran(h)$ (i.e. all events mentioned in $\phi$ are in the range of $h$) then $trans_h(\phi)$ is the same as $trans_{h_{tot}}(\phi)$.
\end{lemma}

\noindent {\bf Proof} This is because $h^{-1}$ and $h_{tot}^{-1}$ are the same for all events in $ran(h)$.
\hfill $\Box$\\


The following lemma provides a result that allows us to demonstrate the relationship between an infinite trace $u$ preserving a transformed temporal property and a corresponding transformation on the trace preserving a temporal property.
\begin{lemma}
\label{lemma:pathModelsPhi_extended}
If $\alpha(\phi) \subseteq ran(h)$, then \\
$u \in (dom(h))^* \Rightarrow \; (u \models trans_h(\phi)  \Leftrightarrow  h(u) \models \phi) $
\end{lemma}
\noindent{\bf Proof:}
By structural induction over the structure of $\phi$.  \\
{\it Base Case:} Consider when $\phi$ is $[x]$. If
\vspace*{-0.2cm}
\begin{eqnarray*}
u & \models & trans_h[x] \\
&\Leftrightarrow& u \models \underset{y \mid h(y) = x}{\bigvee}y \qquad \qquad \qquad \quad \thickspace  [by \; Definition~\ref{defn:trans}]\\
&\Leftrightarrow&  \underset{y \mid h(y) = x}{\bigvee} u \models [y] \qquad\qquad \qquad \quad [by \; model \; distribution] \\
&\Leftrightarrow& \underset{y \mid h(y) = x}{\bigvee} u_0 = y \qquad\qquad \qquad \quad \thickspace \thinspace [where\; u_0 \; is \; first \; element\; of \; u]\\
&\Leftrightarrow& \underset{y \mid h(y) = x}{\bigvee} (h(u))_0 = h(y) = x   \qquad [condition \; on \; y] \\
&\Leftrightarrow& (h(u))_0 = x  \Leftrightarrow h(u) \models [x]
\end{eqnarray*}

\noindent {\it Inductive Cases:}
Consider where $\phi$ is $\neg \phi$.  Then

$u \models trans_h(\phi) \Leftrightarrow u \not \models trans_h(\phi) \Leftrightarrow  h(u) \not \models \phi  \Leftrightarrow h(u) \models \neg \phi $. \\
Proof of the other cases follow similarly.
\hfill $\Box$\\


The next lemma allows us to demonstrate the relationship between a property $\phi$ which is $\beta$-dependent following transformation.
\begin{lemma}
\label{defn:bDependent_extended}
If $\alpha(\phi) \subseteq \beta \subseteq ran(h)$, then if $\phi$ is $\beta$-dependent then $trans_h(\phi)$ is $h^{-1}(\beta)-$dependent
\end{lemma}
\noindent{\bf Proof:} 
Consider an infinite trace $v$. Then 
\begin{eqnarray*}
 v &\models & trans_h(\phi)\\
& \Leftrightarrow & v \models trans_{h_{tot}}(\phi) \qquad \qquad \quad  \thinspace [by \; Lemma~\ref{lem:h1} \; since \; \alpha(\phi) \subseteq \beta \subseteq ran(h)] \\
& \Leftrightarrow & h_{tot}(v) \models \phi \qquad \qquad\qquad\quad  \thinspace [by\; Lemma~\ref{lemma:pathModelsPhi_extended}] \\
&\Leftrightarrow & h_{tot}(v) \project \beta \models \phi \qquad \qquad \quad \thickspace  \thickspace [by \; Definition~\ref{defn:bDependent}] \\
&\Leftrightarrow & h_{tot}(v \project  h_{tot}^{-1}(\beta)) \models \phi \qquad \quad  \thinspace [by \; function \; mapping] \\
&\Leftrightarrow &  v \project  h_{tot}^{-1}(\beta) \models trans_{h_{tot}}(\phi)  \quad  \thickspace \thinspace [by \; Lemma~\ref{lemma:pathModelsPhi_extended}] \\
&\Leftrightarrow &  v \project  h^{-1}(\beta) \models trans_h(\phi) \qquad \thickspace [since \; \alpha(\phi) \subseteq A and \beta \subseteq A, so \; h \; and \;h_{tot}  \\
&& \qquad \qquad\qquad\qquad \qquad \qquad\qquad\qquad \qquad \qquad\qquad have \;the \; same \; effect]
\end{eqnarray*}

 \hfill $\Box$

\noindent{\bf Proof of Lemma~\ref{lemma:chain_ltlpreserve_extended}}
Let  $u$ to be an infinite execution of $M_n$ since $M_n$ is deadlock free.   
Let $h=g_{i,n}$.  It follows from Theorem 6.5 of~\cite{DBLP:journals/fac/SchneiderTW14} that  $h_{tot}(u) \project \alpha(M_i)$ is an infinite execution of $M_i$ 
Then we obtain 

\begin{tabular}{llll}
$h_{tot}(u) \project \alpha(M_i) \models \phi$  & [since $M_i \models \phi$] \\
$\quad \implies \quad h_{tot}(u) \project \alpha(M_i) \project \beta \models \phi$  & [since $\phi$ is $\beta$-dependent] \\
$\quad \implies \quad  h_{tot}(u) \project \beta \models \phi$ & [$\beta \subseteq \alpha(M_i)$] \\
$\quad \implies \quad  h_{tot}(u \project h_{tot}^{-1}(\beta)) \models \phi$ & [interaction of projection and mapping]  \\
$\quad \implies \quad  u \project h_{tot}^{-1}(\beta) \models trans_{h_{tot}}(\phi)$  & [by Lemma~\ref{lemma:pathModelsPhi_extended}] \\
$\quad \implies \quad  u \project h^{-1}(\beta) \models trans_{h}(\phi)$ & [by Lemma~\ref{lem:h1}] \\
$\quad \implies \quad  u \models trans_h(\phi)$ & [by Lemma~\ref{defn:bDependent_extended}]
\end{tabular} \\
as required.  \hfill $\Box$

\end{document}